\documentclass[11pt,a4paper]{article}
\usepackage{amsmath}
\usepackage{latexsym}
\usepackage{theorem} 
\newtheorem{teorema}{Theorem}[section] 
\newtheorem{definicion}[teorema]{Definition}
\include{amslatex}
\newtheorem{proposicion}[teorema]{Proposition}

\newtheorem{corolario}[teorema]{Corollary}

{\theorembodyfont{\rmfamily} } 
{\theorembodyfont{\rmfamily} \newtheorem{ejemplo}[teorema]{Example}}

\numberwithin{equation}{section}

\include{amsmath}

\begin{document}
\title{{\Large {\bf A Finslerian version of 't Hooft Deterministic Quantum Models}}}
\maketitle

\author{
\begin{center}Ricardo Gallego Torrome\\[3pt]

\end{center}}

\begin{abstract}
  Using the Finsler structure living in the phase space associated to the
tangent bundle of the configuration manifold, deterministic models at the
Planck scale are obtained. The Hamiltonian function are constructed directly from the 
geometric data and some assumptions concerning time inversion symmetry. The existence of a 
maximal acceleration and speed is proved for Finslerian deterministic models. We investigate 
the spontaneous symmetry breaking of the orthogonal symmetry $SO(6N)$ of the Hamiltonian of 
a deterministic system. This symmetry break implies the non-validity of the argument used to 
obtain Bell's inequalities for spin states. It is introduced and motivated in the context of 
Randers spaces an example of simple 't Hooft model with interactions.
\end{abstract}  

\section{Introduction}

The possibility to use deterministic models 
at the Planck scale has been presented for instance in references [1] and [2]. Following 
these ideas, Hilbert space techniques are
useful tools to deal with probabilistic predictions at atomic, nuclear or Standard Model 
scale physics. Quantum mechanics is considered to be a powerful formalism to deal with the 
chaotic evolution of these systems. However the behavior of physical systems at the Planck 
scale can be very different. Therefore deterministic models can be useful to describe the 
physical systems at this more fundamental level.

Particular motivations to investigate deterministic models at the Planck scale have been 
explained in [1]. We recall briefly some of these reasons: 
\begin{enumerate}
\item Firstly, there is the feeling that fundamental concepts like locality, space and time 
are becoming more and more obscure in Contemporary Physics and that this tendency will 
gradually grow in modern quantum theories. It seems that it is not a nice consequence of 
Modern Physics.

\item Secondly, there are conceptual problems related with quantum cosmology. Let us suppose 
that the physical system being described is the entire universe through a master quantum 
wave function. The meaning of this wave function of the universe is problematic because we 
can not make any experiment to test the correctness of it: we live in one universe only and 
we do not have an ensemble of identical universes to check the probabilistic predictions of 
the theory. It seems it is not possible to contrast a quantum model of the whole universe.

\item Thirdly, black hole Physics is problematic from the point of view of Quantum 
Mechanics. The research in this area has produced, among other results, the discovery of a 
fundamental principle as the holographic principle [2]. The interpretation of this principle 
is not intuitive from a field theory point of view; let us consider the fundamental area

\begin{displaymath}
A_p =4 \, ln2\, L_p ^2, 
\end{displaymath}
where $L_p $ is the Planck length. This principle can be stated the following way
\paragraph{}
{\it The quantum degrees of freedom at the Planck scale of a physical system are distributed 
on a surface such that corresponds one boolean degree of freedom for fundamental area 
$A_p$.}
\paragraph{}
In a local quantum field theory the density of states is proportional to the volume of the 
system. Therefore, an interpretation of the holographic principle in the framework of a 
local quantum field theory becomes difficult. 

\item Finally, physicists have found strong difficulties in their attempts to unify quantum 
mechanics with a theory of gravity. This unified theory should be important at short 
distances, where gravity become strong and comparable to other interactions. Usually the 
strategy consist on searching for the right laws of gravity at short distances, maintaining 
Quantum Mechanics as a complete theory. This persistent problem is a reason to reconsider 
the status of Quantum Mechanics as a fundamental theory at the scale where is gravity so 
strong enough as other interactions. 
\end{enumerate}

The approach advocated in reference [1] is to investigate deterministic systems at the 
Planck scale as an alternative way to solve these problems. Due to a dissipative dynamics, 
after a long term evolution, different states evolve to the same one, reducing the 
dimensionality of the Hilbert space. All the ontological states evolving to the same state 
define an equivalence class. The equivalence classes can be non-locally defined and it is 
speculated they are the states described by rays of a Hilbert space, as they are represented 
in Quantum Mechanics.
In addition it was showed that the use of Hilbert space theory in the description of these 
deterministic models is useful to find the connection with ordinary Quantum Mechanics.

The Hilbert space approach to deterministic systems has some problems. The main one is that 
the Hamiltonian of a deterministic system is linear in the momentum variables and therefore 
is not bounded from below. This implies the instability of the system. In addition, only few 
examples are known with a mechanism generating a Hamiltonian bounded from below but these 
models do not involve interactions. Moreover, any deterministic description of Quantum 
Mechanics seems to be plagued by Bell's inequalities. It was conjectured in [1] that at the 
Planck scale physical systems do not meet the required symmetries used in the proof of 
Bell's inequalities for spin states. The reason is that at this level the system can be so 
complex that usual rotation symmetries does not hold. 
\paragraph{}
It is known the geometric nature or interpretation of important physical models. For 
instance, the point particle action is the length of a curve, the string action is a 
generalized area and Yang-Mills actions are functional of connections on bundles. General 
Relativity also has an interpretation in terms of Semi-Riemannian geometry. The objective 
nature of the geometric entities (that is, covariance respect a transformation group) 
implies the relevance of the geometric actions, making apparent the independence of the 
physical phenomenon from the particular way of description adopted. We can said that the 
actions of important physical models are related with Riemannian or Semi-Riemannian 
structures an generalized notions of distance metrics.

Finsler structures are actually natural and constitutes a branch from Differential Geometry 
and Theoretical Physics with a huge recent development (some general references for Finsler 
geometry and some of its applications are [4]-[9]). Finsler structures are as natural as 
Riemannian structures but have less restrictions than Riemannian ones and sometimes seems 
strange why they not appeared recently in field theory in a natural way. Nevertheless 
recently considerable work have be done in some applications in String Theory and General 
Relativity (for instance in [6]) and in the Thermodynamics Theory in curved spaces ([6] and 
references therein). Previous applications of Finsler geometry to physics were performed by 
the school of Miron at Rumania ([7]-[9]). 
\paragraph{}
However our application of Finsler geometry to get deterministic models is complete 
different. It is based on the following general ideas: 
\begin{enumerate}
\item In the construction of physical models geometric structures are of extreme importance. 
Indeed more of the main theories like string and General Relativity are formulated using 
metric theories, in particular structures existing in a pseudo-Riemannian manifold. Metric 
structures are usually required and natural.
\item If we are looking for a formalism that could be also applicable to the whole universe, 
it should contain an irreversible element. We live in one universe and the notion of 
reversible law is maybe not completely valid because it is not completely under experimental 
control when it is applied to a large portion of the universe or to the whole universe.
\end{enumerate}

Finsler geometry have enough ingredients to address both points. Probably is not the only 
possibility, but we were able to use this sophisticated geometry to find some results that 
maybe can be useful for future research on this line. Finslerian distance (ussualy 
associated to the length of a curve using the Finsler metric) can be non-symmetric, that is, 
the "distance between the point $a$ and the point $b$ is not the same than the distance from 
$b$ to $a$. We consider that non-symmetric Finsler metrics (that means when the above 
a-symmetry it is possible) are useful to describe the behavior of irreversible evolutions at 
the fundamental scale. For example, the action of a particle moving in a Finsler space is 
not invariant under the inversion of the parameter of the curve. This asymmetry in the 
metric implies the possibility to describe an irreversible evolution from a geometric point 
of view. This is the main reason to use models in Physics based on Finsler geometry (another 
idea of how to describe Thermodynamics using Finsler structures can be found in [5]. However 
our application mainly differs from this one because we are concerning with the most basic 
level in the description of phenomena).

The aim of the present work is both. Firstly, we investigate some general consequences of 
the `t Hooft theory. Secondly, in order to give a geometric basis for the theory, we explain 
the relation of `t Hooft's models with the Finsler structure of the cotangent bundle of the 
tangent bundle of the configuration manifold ${\bf M}$, ${\bf T}^* {\bf TM}$. Finslerian 
models are free of some problems of the initial 't Hooft's theory. In particular, they 
provide a geometric argument to obtain a lower bound for a Hamiltonian coming from a 
deterministic system.
\paragraph{}
The structure of this work is as follows. In {\it Section} 2, the basic notions and results 
of the 't Hooft theory are presented. Also, the main problems of this approach are 
explained. 

In {\it Section} 3, the use of Finsler geometry to find deterministic models is presented. 
In addition we develop some consequences of the finslerian approach as the existence of a 
higher limit for generalized physical acceleration and speed. 
We describe the spontaneous symmetric breaking that can happening in Finslerian 
deterministic model. The possible absence of Bell's inequalities for spin at the Planck 
scale is also argued, but the argument can be changed to a more general framework than 
finslerian models, provide a geometric argument is possible.

In {\it Section} 4 we to discuss our results in the context of the geometry of spaces of 
smooth Finsler structures is made. We describe a simple deterministic model where 
interactions are present. This model is based on some geometric construction and some 
additional physical requirements. 

Finally, we attach in {\it Appendix A} the basic definitions and results of Finsler geometry 
and other geometric objects mentioned in this work.

\section{The 't Hooft Theory} 

Gerard't Hooft has investigated the possibility to use deterministic models in order to 
describe physical systems at the Planck scale through a Hilbert space formulation ([1],[2]) 
of these models. 
The physical system is described by an eigenstate $\mid x> $ of a set of commuting operators 
$\{ \hat{X}_i (t)\}$, 
\begin{displaymath}
[\hat{X}_i(t),\hat{X}_j (\tilde {t})]=0, \forall i,j,\quad \hat{X}_i (t) |x> =x_i (t)|x> ,
\end{displaymath}
such that the eigenvalues $\{x_i (t)\}$ completely describe the state of the system. This 
states are called ontological. The parameter $t$ is associated with a macroscopic phenomenon 
or device and used as the time parameter by an macroscopic observer, although we are 
considering, microscopic processes, that is at the level of the determinisic description. At 
each instant $t$, the physical system is in correspondence with a particular vector defined 
by the set of eigenvalues $\{ x_i (t) \}$. This set of functions define the real 
configuration of the system at any instant $t$. The Hilbert space is generated by the 
vectors representing the configurations of the physical system. A linear combination of 
elements of a basis of the Hilbert space produces a vector such that the square of the 
module of each component is the probability of the system to be in this particular state. 

The Hamiltonian of a deterministic system with $6N$ degrees of freedom in the phase space is   
\begin{equation}
{\bf H}=\sum ^{6N}_{i=1} p_i f^{i}({x})+g(x).
\end{equation} 
$(x,p)$ are canonical variables, $\{x_i ,p_j \}=\delta _{ij}$. After canonical quantization 
this Hamiltonian reproduces the evolution differential equations, which are the Heisenberg 
equations for the operators $\{ X^i ,\, i=1,...,6N\} $,
\begin{equation}
\frac{d\hat{X}^{i}}{dt}=f^{i}(\hat{X}),\quad i=1,...,6N.
\end{equation}  
When we take the average value of the equation $(2.2)$ we obtain 
\begin{displaymath}
<\tilde{x}| \big(\frac{d\hat{X}^{i}}{dt}-f^{i}(\hat{X})\big) | {x}>=0, \quad i=1,...,6N.
\end{displaymath}
This implies the classical ordinary differential equations
\begin{equation}
\frac{dx^{i}}{dt}=f^{i}(x),\quad i=1,...,6N
\end{equation}
because the scalar product of the Hilbert space is positive defined.
Any system whose evolution is given by the equations (2.3) and has a complete and defined 
set of initial conditions is called deterministic.

Let us consider the quantization of the Hamiltonian $(2.1)$. It does not have a minimal 
eigenvalue because it is linear in momentum. However, the existence of a ground state is 
essential for the stability of the physical system. This parameter is a fundamental 
difficulty in the Hilbert space formulation of deterministic systems.

This problem let us consider a dissipative dynamics a system with a rather turbulent or 
chaotic behavior at the beginning can reach stability in a finite time. This kind of 
dissipation implies the possibility to define the physical states as equivalence classes at 
equilibrium. An equivalence class is defined by the set of ontological states that after a 
long term in the parameter $t$, evolve to the same final state. 

't Hooft have proposed the following solution to the problem of the missing of the 
lower-bound of the Hamiltonian:
{\it If dissipation of information it is possible, the final Hamiltonian could be
bounded from below}. 
It was suggested in reference [1] and [2] that the actual Quantum Mechanics describes not 
the basic degrees of freedom of our universe, but the dynamics of equivalence classes 
reached by these basic states after a long term evolution with a dissipation of information: 
various states can evolve to the same equilibrium state.
The ontological states follow a deterministic dynamics
which is described by the set of first order, ordinary differential equations of the type 
$(2.3)$ (in addition with a complete set of initial conditions). These states are locally 
well defined. By contrast, the equivalence classes of states could not be locally well 
defined and their evolution is quantum mechanical.

This evolution onto equivalence classes can solve the problem of the ground state because
the number of them is smaller than the number of
ontological states. It could be that even with an infinite number of
ontological states, we have a finite number of equivalence classes, a finite
Hilbert space and as a consequence the Hamiltonian has a defined ground state ([1]). Several 
examples has been found by 't Hooft where there is a mechanism producing a Hamiltonian with 
lower bound: the free bosonic system, the  free Maxwell field and the free mass-less 
neutrino system are deterministic systems. These examples at least prove the existence of 
deterministic models with Hamiltonian bounded from below.
 
In this work we denote by a 't Hooft model a deterministic system with a mechanism producing 
a lower bound for the final Hamiltonian at equilibrium. By dissipative mechanism we mean an 
information loss mechanism.

\section{Finslerian Deterministic Quantum models at the Planck scale} 
Let us denote by ${\bf M}$ the configuration manifold of all the degrees of freedom at the 
Planck scale. By configuration manifold we mean a sub-manifold ${\bf M}$ of a $2n$-manifold 
${\bf N}$ such that ${\bf TM}={\bf N}$. With this definition we adopt the formalism of 
Lagrange spaces ([7]) (indeed dual of Lagrange spaces), instead of considering the formalism 
of higher order Lagrange spaces([8]).

The relation between Finsler structures and deterministic systems is based on the following 
points: 
\begin{enumerate}
\item The ontological states at the Planck scale are described by points of the phase space 
${\bf T^* TM}$ and the tangent bundle ${\bf TM}$ is equipped with a dual Randers metric 
$F^*$.
\item The reduction of the
ontological Hilbert space to the quantum mechanical Hilbert space is in correspondence with 
the reduction of the Randers structure $({\bf TM},F^* )$ to the Riemannian structure $({\bf 
TM},h )$. We postulate that this reduction corresponds to the average operation investigated 
in reference [10].
\item We assume that for each particle with generalized velocity $y=\frac{dx^i }{dt}$ there 
is another particle associated such that it is evolving backwards
in time $t$ with velocity $-y$ and the separation between them is zero. 
\end{enumerate}
For the definition of a Finsler structure and Randers structures, basic notions in the 
present work, we refer to the {\it Appendix A} or to reference [4]. The term dual makes 
reference to the manifold ${\bf T^* TM}$ where $F^* $ lives in our formalism.
This spaces are treated in the literature (see reference [6] and references therein) and are 
called Cartan spaces. Points $1$ and $2$ are the main link between Geometry and Physics in 
our proposal. Point $1$ is a nominative axiom, relating notions from Geometry and Physics. 
Point $2$ refers to the link between the geometric theory of Finsler geometry described in 
[10] with a reduction of the associated Hilbert space. This point is capital to find the 
Link between Quantum Mechanics and Deterministic Models at the Planck scale. It is only 
explanatory; while point $1$ is completely arbitrary, point $2$ is just a consequence of 
point $1$ and the theory developed in [10]. Point $3$ is a generalization of the 
particle-anti-particle creation in the context of a non-symmetric geometry background. Due 
to this asymmetry, a non-trivial system arise with a defined fundamental time-arrow.

In addition to the above statements we note other two implicit facts in our construction:
\begin{enumerate}
\item There is a microscopic time arrow associated with the mechanism that produces the 
evolution from the Randers structure $({\bf TM},F^*)$ to the actual Riemannian structure 
$({\bf T M},h)$.
\item There is a Hamiltonian function obtained directly from the geometric data contained in 
the Randers structure $({\bf TM},F^* )$:
\end{enumerate}
Consider a Randers function $F^* $ with the following form (see {\it Appendix A} for the 
definition of Randers space),
\begin{displaymath}
F^* (x,p)=\alpha(x,p)+\beta(x,p).
\end{displaymath}
Then, we perform the following identification between the Hamiltonian function and the 
non-symmetric part of the Randers function,

\begin{equation} 
{\bf H}=\sum^{n}_{i=1}
p_{i}f^{i}({x})\longrightarrow 2\sum^{n}_{i=1} {\beta}^{i}(x)p_{i}
\end{equation}
and if we identify component by component,
\begin{equation}
 2{\beta}^{i}=f^{i},\quad i=1,...,6N.
\end{equation}
The ordinary differential equations $(2.3)$ are then
\begin{equation}
f^{i}={\beta}^{i}=\frac{dx^{i}}{dt},\quad i=1,..,6N.
\end{equation}
In order to quantize the model we use the canonical quantization through the prescription
\begin{equation}
x^i\longrightarrow \hat{X}^i,
\quad
\beta^i (x)\longrightarrow \beta^i (\hat{X}),
\quad
p_{i}\longrightarrow -i\frac{\partial}{\partial x^{i}}=\hat{P}_i.
\end{equation}
This representation holds the canonical quantization relation:
\begin{displaymath} 
 [\hat{X}_i ,\hat{P}_j ]=\delta
_{ij} .
\end{displaymath}

The reason why we choose the above Hamiltonian function $(3.1)$ is the following: the first 
term corresponds to a particle moving forward in time while the second term corresponds to a 
particle moving backward in time, both at the same position; there is a democracy in the 
choice of the macroscopic time arrow. 

We would like to justify more in detail the Hamiltonian function $(3.1)$. Since we use phase 
space variables, we translate the above assumption from velocities to momentum variables. An 
example of a dual Finsler structure with Finsler function $F^* $ living in the cotangent 
bundle ${\bf T^* (TM)}$ is defined using the following procedure: if $({\bf TM},F)$ is a 
Finsler structure, let us consider the dual Finsler structure defined by:
\begin{displaymath}
F^* (x,p):=F(x,y_p)\,\,\, \textrm{such that}\,\,\,  y_p(\tilde{p}):=g_{y_p} (p,\tilde{p}),
\end{displaymath}
 $\vec{p}=p^i \frac{\partial}{\partial x^i}\in {\bf T}^* _u {\bf T}_x {\bf M};\, y\in {\bf 
T}_u {\bf T}_x {\bf M}$. $y_p $ is the dual vector of the 1-form $p$ defined by the second 
relation. $g_{y_p}$ is the fundamental tensor of the structure $({\bf TM},F)$ evaluated at 
the point $y_p $ (for the definition, see {\it appendix A}).

The classical Hamiltonian function $(3.2)$ coincides with:
\begin{equation}
{\bf H}=F^* (x,p)-F^*(\tau (x),\tau (p))=2{\beta}^{i}p_{i}.
\end{equation}
The transformation $\tau$ is the time inversion operator respect the microscopic time $t$. 
The action of time inversion operator in the canonical variables is defined such that the 
canonical relation $\{x_i ,p_j \}=\delta _{ij}$ remains invariant.

The quantization of the above models is equivalent to the quantum mechanical description of 
a deterministic system. The quantized Hamiltonian is defined by:
\begin{displaymath}
{\bf 
\hat{H}}=F^*(\hat{X},\hat{P})-F^*(\hat{T}\hat{X}\hat{T}^{-1},\hat{T}\hat{P}\hat{T}^{-1}).
\end{displaymath}
$\hat{T}$ is the time inversion operator. The Hamiltonian is 
\begin{equation}
{\bf \hat{H}}=2{\beta}^{i}(\hat{X})\hat{P}_{i}.
\end{equation}
A simple calculation shows that for this Hamiltonian the relation $\hat{T}{\bf 
\hat{H}}\hat{T}^{-1}=-{\bf \hat{H}}$ holds and that the elementary evolution operator 
\begin{displaymath}
{\bf \hat{U}}(t,t+\delta t)=\hat{I}-i\delta t {\bf \hat{H}}
\end{displaymath}
is invariant under time inversion $\hat{T}$, producing a geometric time arrow (note that 
states are invariant by the time inversion operation).

The Hamiltonian $(3.6)$ is not bounded from below. In order to solve this problem we propose 
the following mechanism: let us define the average classical Hamiltonian defined by
\begin{displaymath}
<{\bf H}>:=\int _{{\bf I}^* _x} {\bf {H}}(x,p)|{\psi}(x,p)| ^2 d^{6N-1}p.
\end{displaymath}
The manifold ${\bf I}^* _x \subset {\bf T^* _x (TM)}$ is defined by ${\bf I}^* _x :=\{ p\in 
{\bf T^* _x (TM)}\mid F^* (x,p)=1 \}$. $|{\psi}(x,p)| ^2$ is a weight function on the 
indicatrix ${\bf I}^* _x $ and it is determined by the geometric data $({\bf TM},F^*)$.

The justification of this construction is the following.
In reference [10] it was proved the existence of a map from the Finsler category to the 
Riemannian category relating the most important geometric notions. This map was basically 
interpreted as ``average" of the finslerian objects (although in [10] we formulated our 
mathematical construction mainly related with the so-named Chern's connection. However, it 
was noted that a similar average operation is also applicable to other connection as the 
Cartan connection or in general to any linear connection in $\pi^*{\bf TM}$. We also think 
that the construction is also extendable to the non-linear connection, a very important 
notion in Finsler geometry). Here we remark that this average operation is also applicable 
to the Hamiltonian operator after canonical quantization of the classical Hamiltonian 
because it is constructed using the Finsler function. This average is interpreted as a long 
term evolution of the initial Hamiltonian. Another more physical reason to integrate only 
over ${\bf I}^* _x$ is the ``Holographic Principle in phase space": all the quantum 
information is contained in a sub-manifold of dimension $n-1$, in this case the indicatrix 
${\bf I}^* _x$. This holographic principle is formulated in the phase space instead of the 
normal formulation in the configuration space and appears as a re-interpretation of the 
positive homogeneity requirement. It seems to be possible to translate this construction to 
the configuration space just through a generalized Fourier transformation.

The above average Hamiltonian function has an associated quantum operator (better a density 
operator) $<{\bf \hat {H}}>$. This operator is defined by
the action on an arbitrary element of the Hilbert space of the states of defined generalized 
coordinates: 
\begin{displaymath}
\hat{<{\bf H}>}_x(\hat{X},\hat{P})\mid p>:=\int _{{\bf I}^* _x} {\bf 
\hat{H}}(\hat{X},\hat{P})|{\psi}(x,p)\mid ^2 |p>d^{6N-1}p =
\end{displaymath}
\begin{equation}
\int _{{\bf I}^* _x} ({\bf H}(x,p)|{\psi}(x,p)\mid ^2) |p+G(x)>d^{6N-1}p,\quad \forall \,\, 
|p>\,\in \mathcal{H}.
\end{equation}
The average quantum Hamiltonian density operator $<{\bf \hat{H}}>(\hat{X},\hat{P})$ is 
linear. $\{\mid p> \}$ is the set of vectors such that the Finsler norm is 1: $\hat{P}^i 
\mid p> =p^i \mid p> $ with $F^* (x,p)=1$. The function $G(x)$ is the translation produced 
by the operators $\hat{X}^i $ on the momentum state $\mid p>$, computable from the canonical 
conditions and the form of the operators $\beta ^i (\hat{X})$.

The first property of the above Hamiltonian $(3.7)$comes from the definition of Randers 
space. All the terms are bounded and positive defined because the functions $\{ \beta ^i \} 
$ are bounded and also because we are integrating only over the indicatrix ${\bf I}^* _x$. 
Therefore we obtain the following result,
\begin{teorema}
Let $({\bf TM},F^* )$ be a Randers space. Then there is a deterministic system with the 
average Hamiltonian density defined by the relation (3.7) and with ${\bf \hat{H}}$ defined 
by the {\it equation} (3.6). Then the average Hamiltonian $<{\bf {\bf H}}>_x$ is bounded. 
\end{teorema}

The local converse of this result also holds, proving the generality of the connection 
between deterministic systems and Randers geometry,
\begin{teorema}
Let ${\bf \hat{H}}=2{\beta}^{i}(\hat{X})\hat{P}_{i}$ be a quantum Hamiltonian operator 
describing a deterministic system. Suppose that the average Hamiltonian is bounded. Then 
there is a Randers structure that reproduces the above Hamiltonian and the Randers function 
is defined locally at one point by the expression: 
\begin{displaymath}
F(x,p)=\sqrt{\delta _{ij}p^i p^j}+f_i p^i .
\end{displaymath} 
\end{teorema}
{\bf Proof:} We read from the Riemannian metric and the 1-form that characterizes the 
Randers structure from the Hamiltonian; the Hamiltonian of a deterministic system is of the 
form ${\bf H}=f^i (x)p_ i$. We associated the following structure locally such at the point 
$x$ it is given by:
\begin{displaymath}
a_{ij}=\delta_{ij},\quad 2\beta ^i (x)=f^i (x),
\end{displaymath}
where the functions $f^i (x)$ characterize the deterministic system. That the final 
Hamiltonian is bounded implies the functions $\beta ^i (x)$ are also bounded, that is a 
fundamental requirement to obtain a Randers function.\hfill $\Box$

It is important to note that the Randers structure of the thesis of {\it Theorem 3.2} and 
the Randers structure of the hypothesis of {\it Theorem 3.1} are not the same. The reason is 
because they describe different deterministic systems. In addition let us note that the 
Riemannian structure $a_{ij}=\delta_{ij}$ is arbitrary: our choice was the simplest one, but 
it can be constrained because the topology of the manifold ${\bf TM}$ (although locally the 
Finsler structure looks like in {\it theorem $3.2$}) and by physical consistency as we will 
see ($\beta$ should be bounded by $a_{ij}$). When the metric $a$ is given globally we can 
extend $beta$, defining a Randers space in the whole manifold ${\bf TM}$, assuming that it 
is simply connected.
\paragraph{}
Another consequence of the geometric origin of the Hamiltonian is that because the 
requirements that $F^* $ is a Randers function, the functions $\{\beta ^i \}$ are bounded. 
This implies that generalized velocities and accelerations of the particles are bounded,
\begin{corolario}
Consider a deterministic system associated with a Randers space. Then the generalized speed 
and acceleration of any physical sub-system are bounded.
\end{corolario}
This consequence is interesting because it means the following for our model. From the 
geometric point of view bounded means respect the metric structure $a_{ij}$ of the given 
Randers structure $(\alpha , \beta)$. Because both accelerations and velocities can become 
observables in some models, also happens with the metric $a_{ij}$, which becomes not 
arbitrary and dynamical. The problem to find the correct structure such that the $1$-form 
${\beta}$ is bounded with the Riemannian metric $a_{i j}$ implies some additional hypothesis 
concerning the dynamical behavior of both.

As consequence of the existence of a maximal physical acceleration, there is a limit for the 
strength of the gravitational field, if the strong equivalence principle holds. Therefore we 
are dealing with a theory that contains a finite gravitational interaction.

A simple mechanical model can give an estimation of the value of the maximal acceleration. 
Suppose that the universe has a limited energy content, there is a minimal distance $L_p $, 
the maximal speed is $c$ and the ontological degrees of freedom of the model describe the 
molecules of a classical gas. We can write the elementary work that the rest of the universe 
can make on a defined subsystem. Since this maximal work is equivalent to the energy of the 
particles involved, we obtain the relation 
\begin{displaymath}
L_p m a_p \sim \delta {m}c^2 .
\end{displaymath}
The maximal exchange of energy is bounded by $\sim M_U c^2 $, where $M_U $ is the equivalent 
mass of the total energy of the universe, excluding the sub-system considered.
The mass $m$ appearing in the left side is just the mass of the particle, and if this mass 
is the Planck mass $M_p$, then 
\begin{displaymath}
a_p \sim \frac{M_U }{M_p }\frac{c^2 }{L_p }.
\end{displaymath}
This acceleration is very huge when $L_P$ is the Planck scale,
\begin{displaymath}
a_p \sim \frac{M_U }{M_p }10^{52} m/s^2 .
\end{displaymath}

If the change in the state of the sub-system is only produced by the neighborhood of the 
elementary particle, then instead of $M_U $ there is a mass comparable to $m$. Therefore the 
maximal acceleration is
\begin{displaymath}
a_p \sim 10^{52}m/s^2 .
\end{displaymath}
Note that this acceleration is independent of the mass of the particles. This implies that 
the equivalence principle for the maximal acceleration holds in this limit. In addition, 
this example shows the equivalence between the maximal acceleration and a minimal length 
$L_p $, when there is a maximal speed $c$.
\paragraph{}
As another example of application of our geometric formalism, let us consider the 
Hamiltonian describing a deterministic system with $12$ degrees of freedom associated with 
two pairs of particles living in a space ${\bf M}$ of dimension three. The symmetry group of 
the Hamiltonian is contained in the group $O(12)$ because it is the Euclidean product of two 
vectors of a $12$-dimensional space (by associated particles we mean a pair of identical 
particles such that they are at the same position but one is moving forward and the other 
backward on the external time $t$). Let us consider a particular configuration describing a 
system of two correlated pairs of associated particles and their environment. The symmetry 
group for this special configuration contains the group $O(6)\times O(6)\times G$, where the 
first two terms $O(6)$ describe the symmetry related with the two separated pair of 
particles and $G$ determines the symmetry of any other sub-system.
This configuration implies an spontaneous symmetry break of the group $O(6N)$,
\begin{displaymath}
O(6N)\longrightarrow O(6)\times O(6)\times G .
\end{displaymath}
This symmetry break produces Goldstone's bosons that we consider part of the environment. 

Consider the sub-system composed by two correlated pairs. The symmetry of this Hamiltonian 
is $O(6)\times O(6)$. The existence of an internal time $t$ implies the existence of the 
time inversion transformation $\hat{T}$ defined by the action on the generalized canonical 
coordinates (let us recall that the velocity $y$ is also considered as a coordinate in the 
Hilbert space approach to deterministic systems),
\begin{displaymath}
(x,y)\longrightarrow (x,-y).
\end{displaymath}
Invariance of the canonical quantization implies the transformation
\begin{displaymath}
(\hat{P}_x ,\hat{P}_y)\longrightarrow (-\hat{P}_x,\hat{P}_y),
\end{displaymath}
because the time inversion is an anti-unitary transformation on the Hilbert space ([11]). A 
similar transformation for the classical momentum holds. 

We remark that the consistency of this splitting of the cotangent space ${\bf T}^* {\bf TM}$ 
is based on the existence of an additional geometric structure associated with the time 
inversion $\hat{T}$. This additional structure breaks again the symmetry of the Hamiltonian,
\begin{displaymath} 
O(6)\times O(6)\longrightarrow O^2 (3)\times O^2 (3).
\end{displaymath}
Because the physical system is deterministic and has a well defined momentum and generalized 
position values, it is in a particular defined state. The evolution of these states are in 
one to one  correspondence with the 1-form $(\beta^1 _- (1),\beta^2 _- (1),\beta^3 _- 
(1),\beta^1_+ (1),\beta^2 _+ (1),\beta^3 _+ (1))$ describing the evolution of the first pair 
and $(\beta^1 _- (2),\beta^2 _- (2),\beta^3 _- (2),\beta^1_+ (2),\beta^2 _+ (2),\beta^3 _+ 
(2))$ for the second (the notation $\pm$ corresponds to the splitting induced by time 
inversion in $ {\bf T}_x {\bf M}$). But when the system follows the evolution guided by a 
particular value of the above forms the symmetry is again broken,
\begin{displaymath}
  O^2 (3)\times O^2 (3) \longrightarrow  O^2 (2)\times O^2 (2).
\end{displaymath}
The final group $ O^2 (2)\times O^2 (2)$ is because is the biggest group preserving a 
particular deterministic evolution.
 
Therefore it is not possible that the system could hold a non-trivial irreducible 
representation of the rotation group $SO(3)$ consistent with a deterministic evolution: the 
symmetry group for a defined system of two correlated pair of particles at the Planck scale 
is $ O^2 (2)\times O^2 (2)$. This group is not enough to contain the rotation group $SO(3)$
\begin{teorema}
For a deterministic system composed by two correlated, identical pairs of associated 
particles with energies at the Planck scale, there is not a non-trivial irreducible 
representation of the rotation group leaving invariant the deterministic evolution defined 
by particular values of the beta function.
\end{teorema} 
One consequence of this fact is that the ordinary proof of Bell's inequalities for spin does 
not hold at the Planck scale for this system. The reason is that the proof uses the rotation 
symmetry and it does not hold for deterministic systems at this scale. Even the notion of 
spin is not truly defined in this context. Therefore the claim is that Bell's inequalities 
for spin does not hold for Deterministic Finslerian Models.

At ordinary energies the breaking $ O^2 (3)\times O^2 (3) \longrightarrow  O^2 (2)\times O^2 
(2)$ is not given. Only at high energies of order of the Planck scale we can expect this 
break because it means that the system can not decouple from the ambient in a way that 
rotation transformations of the system have sense. However at ordinary scales this 
decoupling have indeed sense and the above symmetry break does not hold. 

This possible absence of was anticipated by 't Hooft and is independent of the nature of the 
model, finslerian or not. Here we remark the geometric character of this phenomenon in the 
case of Finslerian deterministic models.

\section{Discussion }

The relations between the Finsler structure $({\bf TM},F^* )$ and the Riemannian structure 
$({\bf TM},h)$ is described in reference [10] (indeed it was considered the case of a 
general, smooth manifold ${\bf M}$). In addition, it was shown the existence of a map from 
the category of Finsler spaces to the category of Riemannian spaces mapping the Chern 
connection of $F$ (generally, any linear connection living in $\pi ^* {\bf M}$) to a linear 
connection on $\pi^*{\bf TM}$ and the hh-curvature to the curvature of this linear 
connection. These transformations can be interpreted as ``average" operations of the Finsler 
structures and objects. The physical interpretation of these averages is that the Finsler 
structure living in the phase space manifold ${\bf T^* TM}$ evolves after a long term to the 
equilibrium described by the Riemannian structure ${(\bf TM},h)$. This Riemannian structure 
describes the geometry of the phase space when the system of all ontological states reach 
the equilibrium. However, the Hamiltonian describing the evolution of the averaged system 
when the system has evolved after a long term is not the Hamiltonian coming from the 
``average" Finsler structure $({\bf TM},h)$ (that is indeed Riemannian). The reason is 
because these averaged physical systems are not systems of fundamental particles at the 
Planck scale, but could be composite objects like strings. Since they do not feel times so 
small as the Planck time $L_p/c$, the Hamiltonian guiding their dynamics is the average 
Hamiltonian $<\hat{H}>$, not the deterministic Hamiltonian based on the geometric structure, 
Finslerian or Riemannian. 

When the system arrives to the equilibrium the Finsler structure is just the Riemannian 
structure $({\bf TM},h)$. From the definition of the fundamental or ontological Hamiltonian 
(3.6), we obtain in the equilibrium the condition
\begin{displaymath}
{\bf \hat{H}}=0.
\end{displaymath}
The existence of macroscopic matter structures and gravity can be associated with the 
following decomposition:
\begin{displaymath}
{\bf \hat{H}} =<{\bf \hat{H}}> +\delta {\bf \hat{H}} ={\bf \hat{H}}_{\it matter}+{\bf \delta 
\hat{H}}.
\end{displaymath}
If we take the average on each member of this relation, one obtains at equilibrium
\begin{displaymath}
<{\bf \hat{H}} >=<\,<{\bf \hat{H}}_U>\,> +<\delta {\bf \hat{H}}> ={\bf \hat{H}}_{\it 
matter}\,+\,<{\bf \delta \hat{H}}>=0.
\end{displaymath}
We associate $<{\bf \hat{H}}_{U}>={\bf \hat{H}}_{matter}$, $<\delta {\bf \hat{H}}>={\bf 
\hat{H}}_{gravity}$.
Therefore in this model the distinction between matter and gravity appears as result of a 
long term evolution of the ontological states. Also it appears remarkable that while in 
equilibrium appears gravity to compensate matter, at non-equilibrium (that is when the 
structure is Finslerian) there is some kind of pre-gravity interaction, described by the 
Hamiltonian $\delta {\bf \hat{H}}$. The qualitative characteristic of this interaction 
should be study further. Matter seems identical (the particle content seems complete 
identical because the Hamiltonian for matter, before and after average is the same $<{\bf  
\hat{H}}$). This fact implies the universality of our formalism in order to get any quantum 
system from a deterministic model.

Connecting with 't Hooft theory, we describe in a geometric way the projection from an 
ontological state to an equivalence class as follows:
\paragraph{}
{\it The projection after a long term evolution of a deterministic system to the equilibrium 
equivalence class is described by the transformation that average the dual Finsler 
structures living in the manifold ${\bf T^* TM}$.}
\paragraph{}
It is a remarkable consequence of the finslerian 't Hooft models the prediction of the value 
of a maximal acceleration and speed for physical systems. This can be interpreted as the 
requirement of the existence of two Natural constants by geometric consistency. In addition, 
the possible absence of Bell's inequalities for spin is a remarkable prediction for the 
general 't Hooft models: these inequalities are the main obstructions for the construction 
of hidden variables theories. In the present paper we have showed that the absence of Bell's 
inequalities is possible at the Planck scale, where not decoupling of the system with the 
ambient is taken and even the notion of three dimensional rotations become unclear. This 
explanation is simultaneous with the prediction of the existence of pre-gravity interaction 
in a natural way. Therefore, it is possible the construction of hidden variables theories at 
this  energy without the introduction of non-local actions and the mechanisms could be 
promoted by a pre-gravity interaction.

That all the 't Hooft models have a local geometric interpretation in terms of Finsler 
geometry and the geometric origin of a microscopic time arrow obtained from the geometric 
data, are the mayor goals of these models. In addition we can motivate a deterministic model 
based in some construction of a Randers space containing interactions. Unfortunately the 
model is not completely defined by the geometry and physical hypothesis should be 
introduced.

We start reviewing the treatment of 't Hooft of a deterministic system with a dissipative 
dynamics ([1]). The Quantum Hamiltonian is:
\begin{displaymath}
 \hat{{\bf H}}=\vec{p}\cdot \vec{f}(\vec{g}).
\end{displaymath} 
Consider an scalar operator $\rho(\vec{q})$ such that $[\rho (\vec{q}),\hat{{\bf H}}]$=0. 
Then we can perform the following decomposition:
\begin{equation} 
 \hat{{\bf H}}=\hat{{\bf H}}_1\, -\,\hat{{\bf H}}_2 ;
 \end{equation}
 with 
\begin{displaymath}
\hat{{\bf H}}_1 = \frac{1}{4\rho} (\rho^2 +\hat{{\bf H}})^2;\,\,  \hat{{\bf H}}_2 = 
\frac{1}{4\rho} (\rho^2 -\hat{{\bf H}})^2.
\end{displaymath}
This both Hamiltonian commute, $[ \hat{{\bf H}}_1 , \hat{{\bf H}}_2]=0$.

In order to bounded from below the complete Hamiltonian one can introduce the constrain than 
on physical states the following condition holds:
\begin{equation} 
\hat{{\bf H}}|\psi> \rightarrow 0.
\end{equation}
That should be understood as a long term evolution statement: the physical system evolves to 
states obeying condition (4.2). This constraint immediately implies the bound of the 
Hamiltonian:
\begin{displaymath}
\hat{{\bf H}} \rightarrow  \hat{{\bf H}}_1 \rightarrow \rho^2 \geq 0.
\end{displaymath} 

These constrains can be motivated first if we mimic the system in terms of a non-dissipative 
model and where the system correspond to a quantum oscillator, where all the ``orbits" are 
stable such that $\hat{{\bf H}}=\vec{p}\cdot \vec{f}(\vec{g})$ holds and such that $[{\bf 
H}_s , \rho^2]=0$.

The stable orbits are restricted by the condition:
\begin{equation}
 e^{-\hat{{\bf H}}T}|\psi>=|\psi>,
 \end{equation}
where $T$ is the period of the orbit $\rho =1$. This condition, equivalent to the constrain 
(4.2) and implies the limitation of trajectories to stable orbits at equilibrium.
\paragraph{}
Let us compare this construction with a parallel construction using Finslerian Models.
One start directly with a classical Hamiltonian of the form 
\begin{displaymath}
H=F(\vec{q},\vec{p})-F(\vec{q},-\vec{p}).
\end{displaymath}

After canonical quantization, we identify $\hat{{\bf H}}_1 =F(\vec{q},\vec{p})=\alpha 
+\beta$; $\hat{{\bf H}}_2 =F(\vec{q},\vec{p})=\alpha -\beta$. 
 
We propose the constrain $\rho =1$ on physical states. It is equivalent to an average 
operation defined in the following way:
\begin{equation}
h_{ij}=\int _{{\bf S}} g_{ij}(x,y),
\end{equation}
where the integration is done on the sphere ${\bf S^{6N-1}}\subset {\bf T^*}_u {\bf T}_x 
{\bf M}$. 

Taking this average in the underlying geometric structure corresponds to constrain the 
values of the quantum states: after a long term evolution, the physical states arrive to the 
sub-manifold ${\bf S^{6N-1}}$.

Suppose now the system composed by two identical elementary system, being their dynamics 
described by a deterministic Hamiltonian  of the form (4.1). Let us suppose modeled on 
Randers spaces, so their Hamiltonian are determined by $(\alpha _1 ,\beta _1)$ and $(\alpha 
_2 ,\beta _2)$. The $1$-forms $\beta _i,\, i=1,2. $ have norm less than one by the 
corresponding Riemannian norms $\alpha _i,\, i=1,2$. 
There are at least two ways to produce a bigger Randers space using just the above geometric 
data:
\begin{enumerate}
\item The first way is valid for complete general structures
\begin{displaymath}
\alpha =\alpha _1 \oplus \alpha _2 ;\,\, \beta =\beta _1 \oplus \beta _2 .
\end{displaymath}
This construction does not produce interaction terms in the total Hamiltonian. There is a 
priori not relation ${\alpha}_1$ ${\alpha}_2$. 
\item The second form recovers the impossibility for a external observer to differentiate  
between identical particles:
\begin{displaymath}
\vec{p}=\vec{p}_1 \times \vec{0} +\, \,\vec{0}\times \vec{p}_2 ;\vec{\beta}=\vec{\beta}_1 
\times \vec{0} +\, \,\vec{0}\times \vec{\beta}_2 ,
\end{displaymath}
\begin{displaymath}
\alpha =\alpha _1 \oplus \alpha _2;\, \, \alpha _1 =\alpha _2.
\end{displaymath}
The quantum total Hamiltonian is given by: 
\begin{displaymath} 
\vec{\beta}(\vec{p})=\big(\frac{1}{2}\vec{\beta} _1 (\vec{p}_1)+\vec{\beta} _1 
(\vec{p}_2)+\vec{\beta} _2 (\vec{p}_1)+\vec{\beta} _2 (\vec{p}_2)\big).
\end{displaymath}
The mixed terms produce the interaction. The condition $\alpha _1 =\alpha _2$ is to ensure 
that the above construction is a Randers space. 
\end{enumerate}
\paragraph{}
In order to conclude the discussion we would like to discuss our idea with some recent 
applications of Finsler geometry in Physics ([6] and references therein). Our application of 
Finsler geometry and in particular, of Randers spaces, seems new even if it contains 
elements that have been already used in other contests. It was performed considerable work 
on higher order mechanics and generalized Finsler spaces ([7]-[9] and also [6]), but the 
bird of deterministic models at the Planck scale is very recent and the application of 
Finsler geometry presented in this paper is also new. In addition we remark that we are not 
concern at this stage with a field theory for these deterministic systems, but with the 
general formalism that we could use to describe them. Finsler geometry is rather intrincated 
complex with so many natural connections for instance like Chern's or Cartan's connections. 
Also the notion of non-linear connection is of fundamental importance. But we are not 
concern on these important topics in this paper because the construction proposed involves 
only notions at the metric level: our average is a universal procedure, valid for any linear 
connection $\pi^* {\bf TM}$. Further research can be provide a mechanism to select a right 
$d$-connection for a field theory of deterministic degrees of freedom. In addition, some 
additional research is needed to understand the extension of the average operation applied 
to the non-linear connection.

\appendix
\section{Basic Results on Finsler Geometry for Deterministic Systems} \label{basic results}

In this appendix we recall the basic notions of Finsler geometry used in the present work.
The main reference for this appendix is [4]. We present the notions for an arbitrary smooth 
manifold ${\bf M}$.

Let $(x,{\bf  U})$ be a local coordinate system over a point $x\in {\bf M}$, where $x\in 
{\bf U}$ have local coordinates $(x^{1},...,x^{n})$, ${\bf U}\subset {\bf M}$ is an open set 
and ${\bf TM}$ is the tangent bundle. We use Einstein's convention for up and down equal 
indices in this work.

A tangent vector at the point $x\in {\bf M}$ is denoted by 
$y^{i}\frac{{\partial}}{{\partial} x^{i}}\in {\bf T}_x {\bf M},\, y^i \in {\bf R}$. We also 
denote by ${\bf TM}$ the set of sections of the tangent bundle. We can identify the
 point $x$ with its coordinates $(x^{1},...,x^{n})$ and the tangent vector $y\in {\bf T}_x 
{\bf M}$ at $x$ with its
 components $y=(y^1 ,...,y^n )$. Then each local coordinate system $(x,{\bf U})$ induces a 
local coordinate system in ${\bf TM}$ denoted by $(x,y,{\bf U})$ such that 
$y=y^{i}\frac{{\partial}}{{\partial} x^{i}}\in {\bf T}_x{\bf M}$ has local natural 
coordinates $(x^1,...,x^n ,y^1,...,y^n)$.

 Let us denote by ${\bf N}={\bf TM\setminus \{  0\} }.$ The notion of a Finsler structure is 
given through the following definition,
 \begin{definicion}
A Finsler structure $F$ on the manifold 
${\bf M}$ is a non-negative, real function  $F:{\bf TM}\rightarrow [0,\infty [$ such that 
\begin{enumerate} 
\item It is smooth in the split tangent bundle ${\bf N}$.

 \item Positive homogeneity 
holds: $F(x,{\lambda}y)=\lambda F(x,y)$ for every $\lambda >0$.
 
 \item Strong convexity holds: 
the Hessian matrix
 \begin{equation} 
g_{ij}(x,y):
=\frac{1}{2}\frac{{\partial}^2 F^2 (x,y)}{{\partial}y^i {\partial}y^j }
\end{equation} 
is positive definite in ${\bf N}$. 
\end{enumerate}
We also denote by a finsler structure in ${\bf M}$ the pair $({\bf M},F)$.
\end{definicion}
The minimal smoothness requirement for the Finsler structure is $\mathcal{C}^5$ in {\bf N} 
when second Bianchi identities are used; more generally only, $\mathcal{C}^4$ differentiable 
structure is required. The matrix $g_{ij}(x,y)$ is the matrix-components of the fundamental 
tensor $g$. The homogeneity condition can be more strong: $F(x,\lambda y)=|\lambda |F(x,y)$. 
Then $({\bf M},F)$ is called absolutely homogeneous Finsler structure.

\begin{ejemplo}
A Randers space is characterized by a Finsler function of the form:
\begin{equation}
F(x,y)=\alpha(x,y)+\beta(x,y),
\end{equation}
where $\alpha (x,y)=a_{ij}(x)y^i y^j$ is a Riemannian metric and $\beta(x,y)=\beta _i 
(x)y^i$. The requirement of being $g_{ij}$ positive definite implies that the 1-form $(\beta 
_1,...,\beta _n )$ is bounded, using the above Riemannian metric $\alpha$. Examples of 
Randers spaces can be found for instance in [4] and [5].
\end{ejemplo}

\begin{definicion}$([4])$
Let $({\bf M}, F)$ be a Finsler structure and $(x,y,{\bf U}) $ a
local coordinate system induced on ${\bf TM}$ from the coordinate system $(x,{\bf U})$ of 
{\bf M}. The Cartan tensor components are
defined by the set of functions: 

\begin{equation}
{ A}_{ijk}=\frac{F}{2} \frac{\partial g_{ij}}{\partial y^{k}},\quad i,j,k=1,...,n.
\end{equation}
\end{definicion}
These coefficients are homogeneous of degree zero in $(y^1 ,...,y^n )$. In the
Riemannian case ${ A}_{ijk}$ are zero and this fact characterizes Riemannian geometry from 
other types of  Finsler geometries (Deicke's theorem).

Since the components of the fundamental and Cartan's tensors have a dependence
on the tangent vector $y$, it is
natural to use other manifold than ${\bf M}$ 
to study Finsler geometry. One possible construction is the following: consider $ {\pi}^* 
{\bf TM}$, the 
pull back bundle of ${\bf TM}$ by the projection  
 
\begin{equation}
{\pi}:{\bf N}\longrightarrow {\bf M}.
\end{equation}
The vector bundle ${\pi}^* {\bf TM}$ has as base manifold ${\bf N}$, the
fiber
over the point $u=(x,y)\in {\bf N}$ is diffeomorphic to ${\bf T}_{x}{\bf M}$ for
every
 point $u\in {\bf N}$ with
 ${\pi}(u)=x$ and the structure group is diffeomorphic to ${\bf GL}(n,{\bf R})$. 

The vector bundle $\pi^* {\bf TM}\subset {\bf TM}\times {\bf N}$ and the projection on the 
first and second factors are given by 
\begin{equation}
\pi _1:\pi ^* {\bf TM}\longrightarrow {\bf N},
\end{equation}
\begin{equation}
\pi _2 :\pi ^* {\bf TM}\longrightarrow {\bf TM}.
\end{equation}
$\pi^* {\bf TM}$ is completely determined as a subset of ${\bf TM}\times {\bf N}$ by the 
following relation; for every $u\in {\bf N} $
and $\xi \in \pi^{-1} _1 (u)$,
\begin{equation}
(u,\xi)\in {\bf \pi^* TM}\quad \textrm{iff} \quad \pi \circ\pi _2(u,\xi )=\pi(u).
\end{equation}
A similar construction ${\bf \pi ^*TM}$ can be performed over ${\bf SM}$, the sphere bundle 
over ${\bf N}$.

One essential notion in Finsler geometry is the non-linear connection. We introduce the 
non-linear connection coefficients, defined by the formula
\begin{displaymath}
\frac{N^{i}_{j}}{F}={\gamma}^{i}_{jk}\frac{y^{k}}{F}-A^{i}_{jk}
{\gamma}^{k}_{rs}\frac{y^{r}}{F}\frac{y^{s}}{F},\quad i,j,k,r,s=1,...,n
\end{displaymath} 
where the {\it formal second kind Christoffel's symbols} ${\gamma}^{i}_{jk}$
are defined in local coordinates by the formula
\begin{displaymath}
 {\gamma}^{i}_{jk}=\frac{1}{2}g^{is}(\frac{\partial g_{sj}}{\partial
x^{k}}-\frac{\partial g_{jk}}{\partial x^{s}}+\frac{\partial g_{sk}}{\partial
x^{j}}),\quad i,j,k=1,...,n;
\end{displaymath}
$A^i _{jk}=g^{il}A_{ljk}$ and $g^{il}g_{lj}=\delta ^i _j .$ Note that the coefficients 
$\frac{N^{i}_{j}}{F}$ are invariant under the scaling $y\rightarrow \lambda y$, $\lambda \in 
{\bf R^+ }$, $y\in {\bf T}_x {\bf M}$.

Let us consider the local coordinate system $(x,y,{\bf U}) $ of the manifold ${\bf TM}$. A 
tangent basis for ${\bf T}_u {\bf N},\, u\in {\bf N}$ is defined by the distributions([4]):
\begin{displaymath}
\{ \frac{{\delta}}{{\delta} x^{1}}|_u ,...,\frac{{\delta}}{{\delta} x^{n}} |_u, 
F\frac{\partial}{\partial y^{1}} |_u,...,F\frac{\partial}{\partial y^{n}} |_u\}, 
\end{displaymath}
\begin{displaymath}
 \frac{{\delta}}{{\delta} x^{j}}|_u =\frac{\partial}{\partial
x^{j}}|_u -N^{i}_{j}\frac{\partial}{\partial y^{i}}|_u ,\quad i,j=1,...,n.
\end{displaymath}

The set of local sections $\{ \frac{{\delta}}{{\delta} x^{1}}|_u 
,...,\frac{{\delta}}{{\delta} x^{n}}|_u,\, u\in \pi^{-1}(x),\, x\in {\bf U} \} $ generates 
the local horizontal distribution $\mathcal{H}_U $ while $\{ \frac{\partial}{\partial 
y^{1}}|_u ,..., \frac{\partial}{\partial y^{n}}|_u ,\,u\in{\pi}^{-1}(x),\, x\in {\bf U} \}$ 
the vertical distribution $\mathcal{V}_U$. The subspaces $\mathcal{V}_u $ and 
$\mathcal{H}_u$ are such that the following splitting of ${\bf T}_u {\bf N}$ holds:
\begin{displaymath}
{\bf T}_u {\bf N}=\mathcal{V}_u \oplus \mathcal{H}_u ,\, \forall \,\, u\in {\bf N}.
\end{displaymath}
This decomposition is invariant by the action of ${\bf GL}(n,{\bf R})$ and it defines a 
non-linear connection (a connection in the sense of Ehresmann) on the principal fiber bundle 
${\bf N}({\bf M},{\bf GL}(n,{\bf R}))$. 

\begin{teorema}(Chern's connection, [4])
Let $({\bf M},F)$ be a Finsler structure. The vector bundle ${\pi}^{*}{\bf TM}$ admits a 
unique linear connection characterized by the connection 1-forms $\{ {\omega}^i _j,\,\, 
i,j=1,...,n \} $ such that the following structure equations hold: 
\begin{enumerate}
\item ``Torsion free" condition,
\begin{equation} 
 d(dx^{i})-dx^{j}\wedge
w^{i}_{j}=0,\quad i,j=1,...,n. 
\end{equation}

\item Almost g-compatibility condition,
\begin{equation} 
dg_{ij}-g_{kj}w^{k}_{i}-g_{ik}w^{k}_{j}=2A_{ijk}\frac{{\delta}y^{k}}{F},\quad i,j,k=1,...,n.
\end{equation}
\end{enumerate}
\end{teorema}
Chern's connection is non-metric compatible but have null torsion. Cartan's connection is 
complete metric compatible, but have torsion. The relation between the Cartan connection 
$1$-forms $(\omega _c )^k _{i}$ in relation with the Chern connection $\omega ^k _{i}$ 
$1$-forms are given by ([4]):
\begin{displaymath}
(\omega _c )^k _{i} =\omega ^k _{i}\, +\, A^k _{i j}\frac{\delta y^j}{F}.
\end{displaymath}  
At this point we should stressed the importance on physical applications of the Cartan 
connection and in general of the metric connections in field theory ([6]). Chern's 
connection, not begin metric compatible is a bit problematic for physical applications, 
being preferible Cartan's connection or d-connection that are metric compatible.
\paragraph{}

The manifold ${\bf I}_x$ is called indicatrix and is defined by 
\begin{displaymath}
{\bf I}_x :=\{y\in {\bf T}_x {\bf M} \mid F(x,y)=1\} .
\end{displaymath}
Let us denote by $\mathcal{F}({\bf I}_x)$ the set of real, smooth functions on the 
indicatrix ${\bf I}_x$. Then the average operation is defined as follows
\begin{definicion}
Let $({\bf M},F)$ be a Finsler structure. Let $f\in {\bf\mathcal{F}}{\bf (I}_x )$ be a real, 
smooth function defined on the indicatrix ${\bf I}_x $ and $(\psi, {\bf I}_x) $ the 
invariant measure. We define the map
\begin{displaymath}
<\cdot >_{\psi}:{\bf \mathcal{F}}{\bf (I}_x )\longrightarrow {\bf R}
\end{displaymath}
\begin{equation}
f(x,y)\longrightarrow \frac{1}{vol({\bf I}_x )}\int_{{\bf I}_x } dvol \, \psi(x,y)\, f(x,y).
\end{equation}
\end{definicion}

In the case of smooth Finsler structures the coefficients $\{ h_{ij},\, i,j=1,..,n \} $ are 
smooth in ${\bf M}$. They are the components of a Riemannian metric in ${\bf M}$,
\begin{proposicion}
Let $({\bf M},F)$ be a Finsler structure. Then the functions 

\begin{equation}
h_{ij}(x):=< g_{ij}(x,y)>,\quad  \forall \,\, x\in {\bf M}
\end{equation}
are the components of a Riemannian metric in ${\bf M}$ such that in a local basis $(x,{\bf 
U})$ is 
\begin{equation}
h(x)=h_{ij} dx^i\otimes dx^j.
\end{equation}
\end{proposicion}

We should mention that the restriction on the indicatrix ${\bf I}_x$ in the integration is 
not necessary: we can perform similar averages procedures on any compact sub-manifold of 
co-dimension $1$ and also of co-dimension $0$ ([10], {\it Proposition 3.13 } and {\it 
Proposition 3.14}). Indeed {\it Proposition 3.14} implies we can take the limit of the whole 
tangent bundle, provided a convenient normalization is used.

The average operation can be extended to obtain average connections and average curvatures 
([10]). This fact can be used to introduce a field theory based on connections as 
fundamental variables for deterministic theories at the Planck scale.

\end{document}